\documentclass[a4paper]{jpconf}
\usepackage{graphicx}
\begin{document}
\title{Dilatonic Brans-Dicke Anisotropic Collapsing Fluid Sphere And
de Broglie Quantum Wave Motion}
% for publication in \jpcs %}

\author{HOSSEIN GHAFFARNEJAD}

\address{Faculty of Physics, Semnan University, Semnan, IRAN, Zip Code: 35131-19111}

\ead{hghafarnejad@yahoo.com; hghafarnejad@profs.semnan.ac.ir}

\begin{abstract}
Two dimensional (2D) analogue of vacuum sector of the Brans Dicke
(BD) gravity [1] is studied to obtain dynamics of anisotropic
spherically symmetric perfect fluid. Our obtained static solutions
behave as dark matter with state equation
$\gamma=\frac{p(\rho)}{\varrho}=-0.25$ but in non-static regimes
behave as regular perfect fluid with barotropic index $\gamma>0.$
Positivity property of total mass of the fluid causes that the BD
parameter to be $\omega>\frac{2}{3}$ and/or $\omega<-1$.
 Locations of the event and apparent horizons of the collapsing fluid are
 obtained in its static regime. In case $\omega>0$ the apparent horizon is covered by
 event horizon where the cosmic censorship hypothesis is still
 valid. According to the model [1], we obtain de Broglie pilot wave of our
metric solution which describes particles ensemble which become
distinguishable via different values of $\omega.$ Incident current
density of particles ensemble  on the  horizons is evaluated which
describe the `Hawking radiation`. The de Brogle -Bohm quantum
potential effect is calculated also on the event (apparent)
horizon which is independent (dependent) to values of $\omega.$
\end{abstract}
\section{Introduction}
Around a collapsing star, gravity takes an ultra-strong intensity.
Hence corresponding matter density and space-time curvature
diverge to infinity. It is predicted from singularity theorems in
general relativity [2] that end of stellar collapse leads to a
visible naked or invisible covered singularity. Property of
visibility of singularities are determined by the causal structure
of collapsing process, satisfying the gravitational field
equations. Internal dynamics of the collapse can be determine type
of singularities. They will be naked and hence visible from view
of external observer if collapse process delays the formation of
the event horizon [3,4,5]. Penrose represented `cosmic
 censorship hypotheses` [6], which explains properties of the final singularity of a
 gravitational collapse. In this conjecture, space-time singularities created from gravitational collapse  should be usually covered
 by the event horizon hyper-surface. Hence
it will be invisible from view of outside observers of collapsed
object namely black hole [7]. However there is still a doubt for
reality of censorship hypotheses and has been
 recognized as one of the most open problems in classical general
 theory of relativity. The censorship hypotheses may to be still valid in quantum gravity approach (see [8] and references therein).
 Black holes and naked
 singularities have different properties observationally. They have different characteristics
 in the gravitational lensing [9,10] and also many exact solutions of Einstein`s gravitational field equations are obtained which
 admit naked singularities
 [11-18].
 In the present paper we use
  vacuum sector of the BD gravity [19] in 2D analogue [1] and obtain non-static solutions of linearized
  field equations.
 Stress tensor of our solutions treats as anisotropic spherically symmetric perfect
 fluid. Equation of state takes dark and visible regular matter regimes during the collapse
 process under particular conditions on different values of the parameters of the system. Apparent and event horizons of the collapsing object
 are
 evaluated where the apparent horizon become inside of the event horizon for $\omega>0$ at static
 regime.
 According to results of the work [1] we obtain de Broglie-Bohm quantum wave functional of the collapsing fluid sphere.
Probability amplitude of particles ensemble and their radial
velocities are calculated from incident current density (ICD) of
particles ensemble. Mass of the collapsing fluid must be takes
some positive real values numerically reaching to conditions
$\omega<-1;\omega>\frac{2}{3}.$ The work is organized as follows:
In section 2 we call dilatonic BD gravity in 2D analogue [1] and
obtain linearized metric solution. Corresponding stress tensor is
obtained to be treat as an anisotropic spherically symmetric BD
perfect fluid. Location of the horizons are determined as
spherically symmetric plan waves. Using some suitable conditions
we obtained time independent horizons location and mass function.
In section 3 we use results of the work [1] to obtain de Brogle
wave functional of the model and then evaluate ICD of particles
ensemble moving on the horizons. de Brogle-Bohm quantum potential
effect is also considered as corrections on our results. Section 4
denotes to summery and conclusion.
 \section{Dilatonic Brans-Dicke collapsing sphere}
One can obtain effective parts of the BD gravity [19] defined on
spherically symmetric curved space times $d s^2=g_{a
b}dx^{a}dx^{b}+\psi^{2}(x^a) \left(d\theta^{2}+\sin^{2}\theta
d\varphi^{2}\right)$ in 2D analogue [1] as $
I[\phi,\psi,g_{ab}]=\frac{1}{2}\int
d^2x\sqrt{g}\{\phi+\frac{1}{2}\phi\psi^2 R+\phi
g^{ab}\partial_a\psi\partial_b\psi+2\psi
g^{ab}\partial_a\psi\partial_b\phi-\frac{\omega\psi^2}{2\phi}g^{ab}\partial_a\phi\partial_b\phi\}$
where $\phi$ and $\omega$  are BD field and parameter
respectively. Experimentally $\omega$ is obtained to be as
$\geq40000$ [20,21]. $\psi^2$ is 2-sphere conformal factor and $R
$ is Ricci scalar of 2D metric $g_{ab}$ defined by 2D coordinates
system $x^{a}$ with $a \equiv 1,2.$ We solved linearized dynamical
field equations and obtained metric solution as
$Gds^2\approx-(1+\varepsilon x_0 e^{a \tau+b\rho})d\tau^2+
2(1+\varepsilon y_0 e^{a\tau+b\rho})d\tau d\rho+(1+\varepsilon
z_0e^{a\tau+b\rho}) d\rho^2 $ where $|\varepsilon|<<1$ is order
parameter of the perturbation series expansion,  $t=\sqrt{G}\tau$
and $r=\sqrt{G}\rho$ (details of calculations are given in ref.
[23]).
 Also we obtained that
$\psi\approx\sqrt{eG}(1+\varepsilon u_0e^{a\tau+b\rho})$ and
$G\phi\approx1+\varepsilon v_0e^{a\tau+b\rho}$ in which $a,b$ are
integral constants and for moving waves with light velocity we
must be set $a=\pm b.$ Other parameters defined by
$\omega,u_0,v_0,y_0,x_0$ and $\gamma$ are obtained exactly against
$a,b$ in ref. [23].  In the next section we show $\varepsilon$ is
related to the corresponding quantum potential of de Brogle wave
evaluated on the event horizon hypersurface. In case of static
regime where $a=0$ we will have $\gamma=-0.25$ describing a dark
matter dominant of the gravitational system and
$b=-\sqrt{\frac{6\omega-4}{\omega+1}},\omega>\frac{2}{3}.$ With
these conditions our metric solutions become asymptotically flat
at large scales $\rho>>1$ for which
$\frac{u_0}{v_0}=\frac{3\omega-2}{5}$,$\frac{x_0}{v_0}=\frac{y_0}{v_0}=\frac{8(3\omega-2)}{5},$
$\frac{z_0}{v_0}=\frac{2(2-3\omega)(2\omega-3)}{5(\omega+1)}.$ In
case of non static regime of the metric solution we must be set
$(a,b)\neq0.$ In the latter case one can obtain moving position of
event horizon by solving $g_{tt}(\tau , \rho)=0$ as $
a\tau+b\rho=-\ln|\varepsilon x_0|$ which in static regime become
$\rho_{EH}=\sqrt{\frac{\omega+1}{6\omega-4}}\ln\left|\frac{4\varepsilon
v_0}{5}(6\omega-4)\right|.$ Surface area of moving apparent
horizon is $4\pi\psi^2(\tau,\rho).$ Its position is obtained by
solving $g^{ab}\partial_a\psi\partial_b\psi=0$ as
$a\tau+b\rho=\ln\left|\frac{2(b^2+2ab-a^2)}{\varepsilon[(a^2+b^2)z_0+2(b^2+ab-a^2)y_0-(a-b)^2x_0]}\right|$
which is
$\rho_{AH}=\sqrt{\frac{\omega+1}{6\omega-4}}\ln\left|\frac{4\varepsilon
v_0}{5}\frac{(3\omega-2)(\omega+5)}{(\omega+1)}\right|$ in static
regime. In the general relativistic limits $\omega\to+\infty$ we
obtain $\rho_{EA}=0.94.$ One can determine mass function $m(r)$ up
to the radius $r$ by calculating $ m(r)=\int_0^r4\pi
r^2\varrho(r)dr$ which by setting $a=0,$ $r=\sqrt{G}\rho,$ and
$b=-\sqrt{\frac{6\omega-4}{\omega+1}},$ we will have
$m^*(\rho)=\frac{m(\rho)}{8\pi\sqrt{G}v_0\varepsilon}=\sqrt{\frac{\omega+1}{6\omega-4}}
\bigg\{1-\frac{1}{2}\bigg[\left(\frac{6\omega-4}{\omega+1}\right)^2-\left(\frac{6\omega-4}{\omega+1}\right)+2\bigg]
e^{-\sqrt{\frac{6\omega-4}{\omega+1}}\rho}\bigg\}$ (see ref.
[23]). Positivity and reality of the mass quantity of the fluid
restrict us to choose $\omega>\frac{2}{3}$ or $\omega<-1$.
\section{de Broglie quantum gravity perspective} According to results of the work [1] (see also [23]) we can obtain phase $\bar{D}$
and amplitude $\bar{R}$ of de Broglie pilot wave
$\Psi(\psi,\phi)=\sqrt{\bar{R}}e^{i\bar{D}}$ of the spherically
symmetric BD action given in the previous section as
$\phi\psi^2=\left(\frac{3+2\omega}{2+\omega}\right)\frac{\bar{R}}{4},$
and
$\psi=\sqrt{G}\bar{R}^{\left(\frac{1+\omega}{3+2\omega}\right)}
e^{i\bar{D}\frac{\sqrt{4+2\omega}}{3+2\omega}}. $ Inserting the
solutions given at the previous section the parts of amplitude
$\bar{R}$ and the phase $\bar{D}$ read $ \bar{R}(\tau,
\rho)=\left(\frac{8+2\omega}{3+2\omega}\right)\exp\{1+\varepsilon
(u_0+v_0)e^{a\tau+b\rho}\}$ and $\bar{D}(\tau,\rho)
=\frac{1}{\sqrt{(4+2\omega)(3+2\omega)}}\bigg\{\frac{1}{2}-(1+\omega)\ln\left(\frac{8+4\omega}{3+2\omega}
\right)+\varepsilon[(2+\omega)u_0-(1+\omega)v_0]e^{a\tau+b\rho}\bigg\}$
which by setting $a=0$ reduces to their static forme (standing
waves) as $\bar{R}(\rho)=\left(\frac{8+2\omega}{3+2\omega}\right)
\exp\{1+\varepsilon
(3v_0/5)(\omega+1)e^{-\sqrt{\frac{6\omega-4}{\omega+1}}\rho}\}$
and
$\bar{D}(\rho)=\frac{1}{\sqrt{(4+2\omega)(3+2\omega)}}\bigg\{\frac{1}{2}-(1+\omega)\ln\left(\frac{8+4\omega}{3+2\omega}
\right)+\varepsilon(v_0/5)(3\omega^2-\omega-9)e^{-\sqrt{\frac{6\omega-4}{\omega+1}}\rho}\bigg\}.
$ We calculate now ICD of particles ensemble containing the Plank
mass $m_p=\frac{1}{\sqrt{G}}$ (in units $\hbar=c=1$) via
$J_{incident}=\frac{i\hbar}{2m_p}\{\Psi\partial_{\rho}\Psi^*-\Psi^*\partial_\rho\Psi\}$
as $
J_{incident}(\rho)=\left(\frac{9-\omega-3\omega^2}{3+2\omega}\right)
\sqrt{\frac{(4+\omega)(3\omega-2)}{(\omega+1)(\omega+2)}}\left(\frac{3\sqrt{2}v_0}{5}\right)
\exp\bigg\{\frac{1}{2}-\sqrt{\frac{6\omega-4}{1+\omega}}\rho+\varepsilon\left(\frac{3v_0}{10}\right)
(1+\omega)e^{-\sqrt{\frac{6\omega-4}{1+\omega}}\rho} \bigg\}.$
  Its value become
$J_{incident}(\rho_{EH})=\left(\frac{9-\omega-3\omega^2}{3+2\omega}\right)
\sqrt{\frac{(4+\omega)}{(\omega+1)(\omega+2)(3\omega-2)}}
\left(\frac{3\sqrt{2}}{8\varepsilon}\right)e^{\frac{5}{8}\left(\frac{3\omega-1}{3\omega-2}\right)}
\bigg\}$ on the event horizon hypersurface and
$J_{incident}(\rho_{AH})=\left(\frac{9-\omega-3\omega^2}{3+2\omega}\right)\sqrt{\frac{(4+\omega)(1+\omega)}{(3\omega-2)(\omega+2)}}
\left(\frac{3\sqrt{2}}{4\varepsilon}\right)
\exp\bigg\{\frac{15\omega^2+58\omega-37}{8(3\omega-2)(\omega+5)}\bigg\}$
on the apparent horizon hypersurface respectively.  Quantum
potential of the de Broglie wave is defined by
$Q=-\frac{\nabla_{\mu}\nabla^{\mu}\sqrt{\bar{D}}}{\sqrt{\bar{D}}}$
which for our metric solutions in static regime become
$Q(\rho)=-\varepsilon^2\left(\frac{3v_0}{5}\right)(3\omega-2)e^{-\sqrt{\frac{6\omega-4}{\omega+1}}\rho}.$
Its value  become $Q(\rho_{EH})=-\frac{3}{8}\varepsilon$ on the
event horizon hypersurface and
  $Q(\rho_{AH})=-\frac{3}{4}\left(\frac{1+\omega}{5+\omega}\right)\varepsilon$ on the apparent horizon
  hypersurface respectively.
 The former result shows relation between quantum potential and perturbation order parameter
 $\varepsilon.$ Namely one can evolve physical meaning of $\varepsilon$ which is supported by quantum potential counterpart
 evaluated on the event horizon as independent of $\omega.$
  Obviously our results and inferences must be
 corrected by regarding backreaction corrections of de Broglie-Bohm quantum potential
 effects on moving particle ensembles (see section 3 in ref. [24]) which is not considered
 here. It is suitable to define relative normalized ICD of
 particle ensemble as
 $\Delta(\omega)=\frac{J_{incident}(\rho_{EH})-J_{incident}(\rho_{AH})}{J_{incident}(\rho_{AH})}=\frac{\exp\{\frac{3\omega}{2(3\omega-2)(5+\omega)}\}}{
 2(1+\omega)}$ and relative normalized quantum potential as $\delta(\omega)=\frac{Q(\rho_{EH})-Q(\rho_{AH})}{Q(\rho_{AH})}=\frac{3-\omega}{2(1+\omega)}.$ We plotted diagrams of $\Delta$ and $\delta$ against $\omega$ with solid and dash lines respectively
 in figure 1. We see potential barrier for $-1<\omega<3$ and potential well  for $\omega<-1 $ and/or $\omega>3$ where one can evolve
 quantum tunneling of particle ensemble! In other words diagram shows that for $-5<\omega<-4.7$
and $-1.6<\omega<-0.8$ we have $\Delta<0$ which means that the
particles ensemble moves into the event horizon.
\section{Concluding remarks} Two dimensional analogue of vacuum
sector of the BD gravity model is used to solve its dynamical
equations. Our classical solutions describe anisotropic perfect
fluid collapsing sphere behaves as dark matter gravitational
source in its static regime and as regular visible matter in its
dynamical non-static regime. Total mass of the fluid is calculated
against the BD parameter for which we must be choose
$\omega\leq-1$ and/or $\omega>\frac{2}{3}.$ In general
relativistic approach $\omega\to+\infty$ ratio of total mass of
the fluid per the Plank mass leads to $M^*=\frac{1}{\sqrt{6}}<<1.$
It is eliminated with $\omega=-1$ and takes infinite value with
$\omega=\frac{2}{3}.$ In short, end of the collapsing fluid
reaches to a black hole structure and covered the causal
singularity satisfying the cosmic censorship hypothesis. de
Broglie-Bohm quantum  pilot wave of the system is also obtained
and corresponding ICD of particles ensemble is evaluated on the
event and the apparent horizons. They describe statistical
approach of thermal `Hawking radiation`. Relative ICD of particles
ensemble moving on the horizons take maximum value where the
quantum potential diverges to an infinite value at $\omega=-1.$

\begin{figure}\begin{center}
\hspace{0cm} \includegraphics[width=5cm]{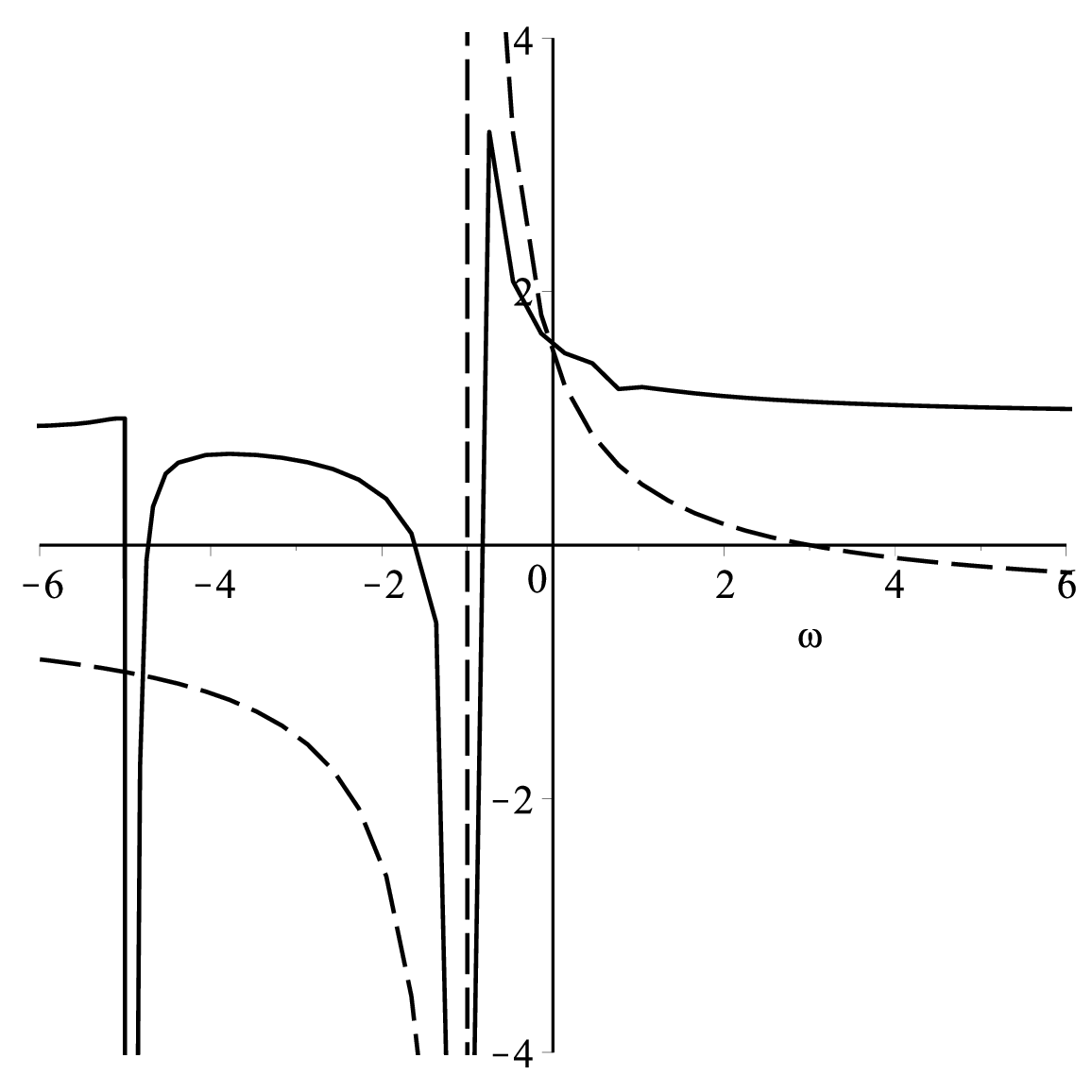} \hspace{0cm}
\caption{\label{mutbeta.figs.} Diagram of $\Delta$ (the normalized
relative ICD of particles ensemble ) and  $\delta$ (normalized
relative quantum potential) is plotted against $\omega$ with solid
and dash lines respectively. }
\end{center}\end{figure}

\end{document}